\documentclass[mathleft,fleqn,%
]{an}
\usepackage{graphicx}
\usepackage[varg]{txfonts}
\usepackage{natbib}
\bibpunct{(}{)}{;}{a}{}{,}

\pdfoutput=1
\begin{document}

\Pagespan{1}{}
\Yearpublication{2014}%
\Yearsubmission{2014}%
\Month{0}%
\Volume{999}%
\Issue{0}%
\DOI{asna.201400000}%

\title{Lessons learnt from the Solar neighbourhood and the Kepler field}

\author{Luca Casagrande\inst{1}\fnmsep\thanks{Corresponding author:
        {luca.casagrande@anu.edu.au}}\fnmsep\thanks{Stromlo Fellow}}
\titlerunning{Photometric and asteroseismic studies of the Galactic disc}
\authorrunning{Luca Casagrande}
\institute{
  Research School of Astronomy \& Astrophysics, Mount Stromlo Observatory, The
  Australian National University, ACT 2611, Australia
}

\received{XXXX}
\accepted{XXXX}
\publonline{XXXX}


\abstract{
  Setting the timeline of the events which shaped the Milky Way disc through
  its 13 billion year old history is one of the major challenges in the theory
  of galaxy formation. Achieving this goal is possible using late-type stars,
  which in virtue of their long lifetimes can be regarded as fossil remnants
  from various epochs of the formation of the Galaxy. There are two main
  paths to reliably age-date late-type stars: astrometric distances for stars
  in the turn-off and subgiant region, or oscillation frequencies along the red
  giant branch. So far, these methods have been applied to large samples of
  stars in the solar neighbourhood, and in the {\it Kepler} field. I review
  these studies, emphasize how they complement each other, and highlight some
  of the constraints they provide for Galactic modelling. I conclude with the
  prospects and synergies that astrometric ({\it Gaia}) and asteroseismic
  space-borne missions reserve to the field of Galactic Archaeology, and
  advocate that survey selection functions should be kept as simple as
  possible, relying on basic observables such as colours and magnitudes only.
}
\maketitle

\section{Introduction}

Late-type stars (broadly FGKM) are long-lived objects and can be regarded as 
snapshots of the 
stellar populations that are formed at different times and places over the 
history of our Galaxy. The fundamental properties of a sizeable number of 
these stars in the Milky Way enable us 
to directly access different phases of its formation and evolution, and for 
obvious reasons, stars in the vicinity of the Sun have been preferred targets 
to this purpose, both in photometric and spectroscopic investigations 
\citep[e.g.,][]{gliese57,wallerstein62,twarog80,stromgren87,edvardsson93,
  nordstrom04,reddy06,c11,hay13,bfo14}. Properties of stars in the solar
neighbourhood, 
in particular ages and metallicities, are still one of the main constraints for 
Galactic chemo(dynamical) models, and provide important clues to understand 
some of the main processes at play in galaxy formation and evolution 
\citep[e.g.,][]{mf89,pcb98,chia01,roskar08,sb09a,jj10,min13,bird13,kpa15}. 

A common feature of all past and current stellar surveys is that, while it is 
relatively straightforward to derive some sort of information on the chemical 
composition of the targets observed (and in many cases even detailed 
abundances), this is not always the case when it comes to stellar masses,
radii, and in particular, ages. When accurate astrometric distances 
are available to allow comparison of stars with isochrones (assuming other 
parameters involved in this comparison -- such as effective temperatures and 
metallicities -- are also well determined), reliable stellar ages can be
derived in restricted regions of the HR diagram, such as the turnoff and
subgiant phase, where stars of different ages occupy clearly different
positions (roughly FG spectral types).
However, even in this favourable condition, statistical techniques are still
needed to avoid biases, in particular that arising from the different
evolutionary speed of stellar models that populate the same region of observed
parameters \citep[the so-called terminal-age bias, e.g.,][]{pe04,jl05,s13}.
\begin{figure*}
\includegraphics[scale=0.088]{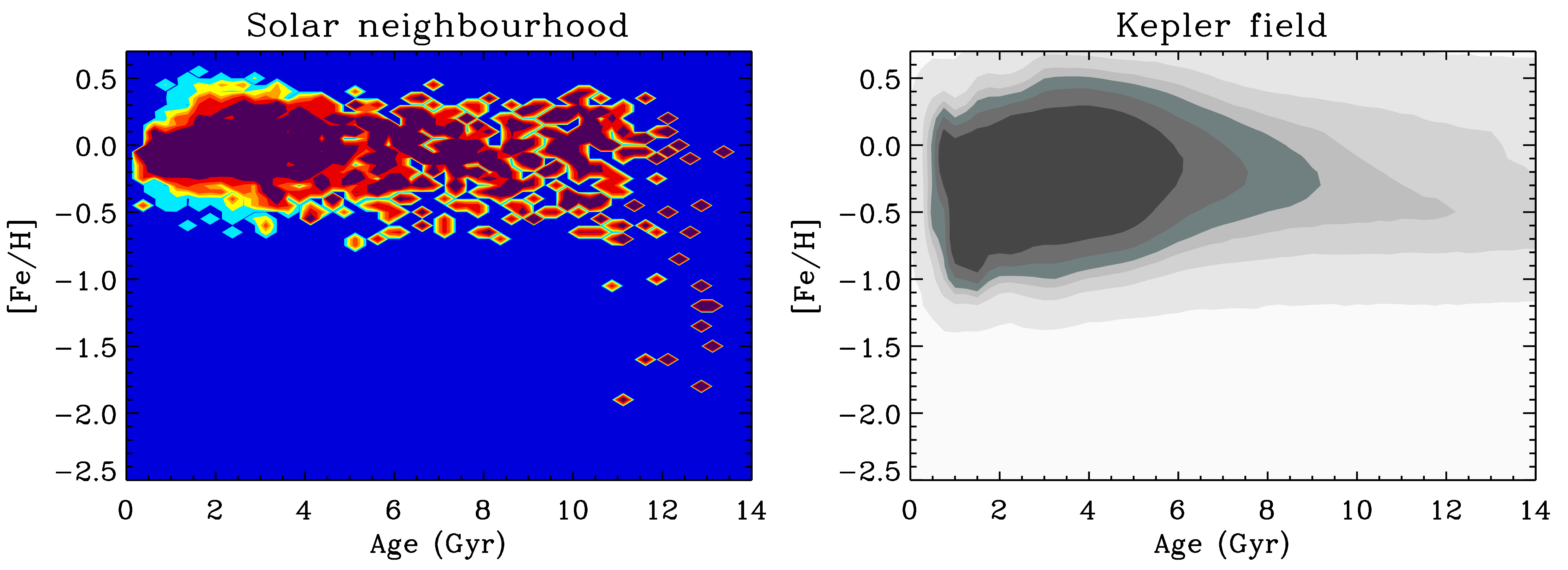}
\caption{{\it Left panel:} age-metallicity relation from the GCS, using stars
  with effective temperatures determined from the InfraRed Flux Method
  ({\it irfm} pedigree), relative age uncertainty below 25\% or absolute
  uncertainty below 1~Gyr (whichever is satisfied). Contours density levels are
  obtained using bins of $0.25$~Gyr and $0.05$~dex, to equally represent
  region with fewer stars. See \cite{c11} for details. {\it Right panel:}
  age-metallicity relation of the {\it Kepler} field from SAGA. Contour levels
  have been obtained by convolving each star with its age, and metallicity
  uncertainties, and assigning a weight proportional to the logarithm of the
  inverse probability of being observed. Notice that sample selection excludes
  metal poor giants, thus preventing us from tracing the early chemical
  enrichment seen e.g. in the left hand panel. See text, and \cite{csv15} for
  details.
}
\label{fig1}
\end{figure*}

The temperature regime of late type stars is also dominated by surface
convection, which is the main driver of the oscillation modes (called
solar-like) that we are now able to detect in several thousands of stars
thanks to space borne asteroseismic missions such as {\it CoRoT} and
{\it Kepler/K2}.
In particular, global oscillation frequencies not only are the easiest ones to 
detect and analyze, but via scaling-relations they are also linked to
fundamental physical quantities such as masses and radii of stars \citep[see
  e.g.,][for a review]{cm13}. Despite the accuracy of seismic
scaling-relations has not yet been fully explored, particularly in the metal
poor regime, stellar radii have been shown to be accurate to better than a few
percent in dwarfs and subgiants 
\citep[e.g.,][]{hub12,vsa12,White:2013bu,coelho}, while masses are likely to be
better than 10\%, but are also less tested \citep{miglio13b}. More
importantly, these relations are applicable to all stars displaying solar-like
oscillations, i.e. also to red giants. Thus, while awaiting for more
stringent tests on scaling-relations, we can already say that asteroseismology
of late-type stars is able to provide stellar masses and radii to an accuracy
that goes from being comparable to, to generally (much) better than achievable
by isochrone fitting.

\section{The Solar neighbourhood}

The region of the Milky Way where we currently have the most complete
chemo-dynamical inventory of stars is the Solar neighbourhood, i.e.~a region
of order 100~pc from us, and for which the {\it Hipparcos} satellite
has measured accurate stellar distances \citep{perryman97,vanLeeuwen07}. The
latter, coupled with proper motions and radial velocities give the complete
six-dimensional phase space information. Fundamental stellar
properties, such as effective temperatures and metallicities (and even
detailed elemental abundances, if possible) provide further pieces to
understand the puzzle of the Milky Way's formation. 

Historically, two different approaches have been adopted to achieve this goal
using stars in the Solar neighbourhood. While spectroscopic studies allow
detailed abundance investigations, they have been limited to small samples of
a few hundred or about a thousand stars at most, and have used sophisticated
kinematic selections to sample significant numbers of members belonging to
different Galactic subpopulations \citep[e.g.,][]{allende04:s4n,reddy06,
  ramirez07,bfo14}. Multi-object spectrographs are now changing this paradigm,
but the combination of instruments' field of view, stellar number densities,
limiting magnitudes etc\ldots makes these facilities better suited to probe
pencil beams through the Galaxy.
Another approach consists instead of using photometry to build large sample of
stars with well defined selection criteria in observational space, typically
colours and magnitudes. In turns, this implies well controlled and/or little
selection biases, though at the price of not being able to derive detailed
elemental abundances from photometry. See e.g., \cite{ibj12} for the latest
review on stellar surveys, or \cite{c15} for the rationale behind photometric
parameters, and a brief discussion of pros and cons between photometric and
spectroscopic surveys.

Arguably, the most complete census of Solar neighbourhood stars is currently
provided by the Geneva-Copenhagen Survey \citep[GCS,][]{nordstrom04} an
all-sky, shallow survey comprising over $14,000$ main-sequence and subgiant
stars closer than $\simeq 100$ pc (40 pc volume limited). The GCS provides the
ideal dataset for studies dealing with Galactic chemical and dynamical
evolution: it is kinematically unbiased, all its stars have radial velocities,
proper motions, {\it Hipparcos} parallaxes, plus highly homogeneous
photometry to derive fundamental stellar parameters. Stellar parameters in the
GCS have undergone a number of revisions \citep{holmberg07,holmberg09,c11}.
All results presented in the following are based on the latest one, which
has improved upon stellar effective temperatures, metallicities and ages.

Given its nearly volume complete nature, the GCS is well suited for the study
of the metallicity distribution function in the solar neighbourhood. The peak
of this function is only slightly subsolar, thus making the Sun a rather
average star given its metallicity. The metallicity distribution function has
been historically used to constrain the gas infall rate \citep[e.g.,][]{lb75},
but slicing it into different age bins suggests that old stars are also a
relevant ingredient in describing the wings of the metallicity distribution
function. A natural explanation to this is provided by radial
migration, where the solar neighbourhood is not only assembled from local
stars, following a local age-metallicity relation, but also from objects
originating from
the inner (more metal-rich) and outer (more metal-poor) Galactic disc that
have migrated to the present position on different timescales
\citep[e.g.,][]{roskar08,sb09a,kpa15}. Continuing along the same line of
reasoning, it thus follows that in the presence of stellar migration, the
age-metallicity relation measured locally will be flat, due to the
superposition of monotonic relations expected at different Galactocentric
radii \citep[see e.g., figure 6 in][]{sb09a}. This is indeed what is seen in
the GCS (Figure \ref{fig1}, left hand panel). As we have already pointed out,
the
GCS has a small spatial extent, but the stellar properties measured locally
can be dynamically
stretched across several kiloparsecs using kinematics. For example, the 
azimuthal velocities ($V$) of stars can be used to derive radial gradients
across several kpc. Likewise, vertical velocities ($W$), or the velocity
dispersion of stars can be use infer vertical gradients well beyond the small
volume covered by the GCS \citep[see e.g.,][]{holmberg07,c11}. In this
context, of particular interests is the age-velocity dispersion relation, a
dynamical tracer of the formation history of the disc. Its exact form has been
debated over the years \citep[e.g.,][]{w77,car85,edvardsson93,bi00,holmberg07,c11}, yet its existence 
clearly indicates the presence of a vertical age gradient in the disc. 
As we see in the next Section, asteroseismology of red giants in the
{\it Kepler} field allows this to be measured quantitatively for the first
time. 
\begin{figure}
\includegraphics[scale=0.74]{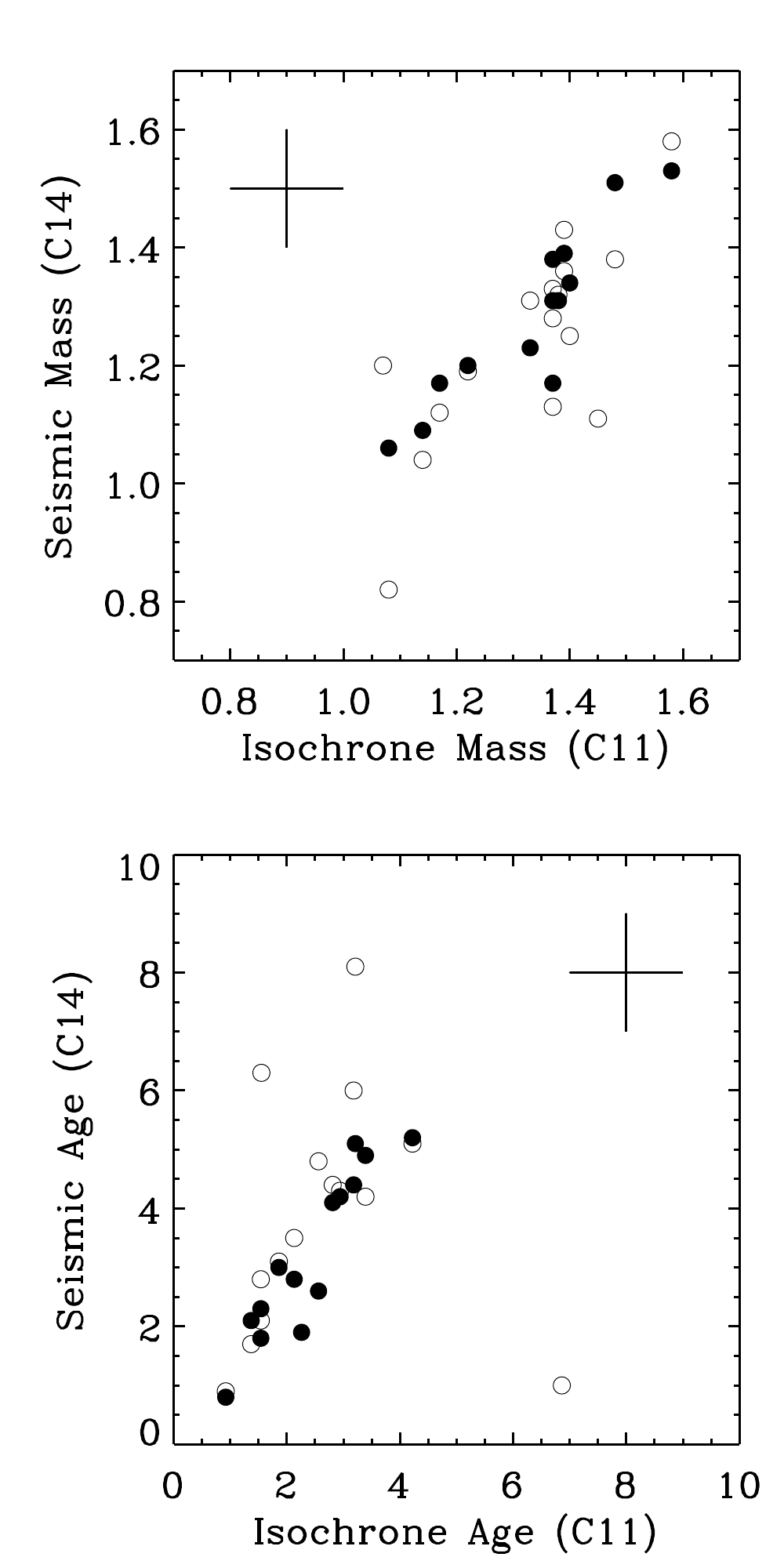}
\caption{{\it Top panel:} comparison between seismic masses \citep{chap14} and
  isochrone masses \citep{c11} for dwarf and subgiant stars in common between
  the GCS and the {\it Kepler} field. Open circles are the seismic results
  obtained when
  assuming an average metallicity of $-0.2$~dex for the {\it Kepler} field
  \citep[see][for details]{chap14}. Filled circles when adopting instead the
  proper [Fe/H] of each star. {\it Bottom panel:} same as above, but
  comparing seismic vs.~isochrone ages. Typical error bars are shown in the
  upper corners.
}
\label{fig2}
\end{figure}

\section{The Kepler field}

Asteroseismology has emerged as a powerful new tool for studying stellar
populations, and the synergy between classical and asteroseismic stellar
parameters is now sought by all major surveys in Galactic archaeology
\citep[e.g.,][]{c14a,pin14,ds15,dc15}. Most of the stars with measured
oscillations in the {\it Kepler} field are red giants, although there are more
than 500 dwarfs/subgiants for which seismic radii, masses and ages have been
derived
\citep{chap14}. Some of these stars are also in the GCS, and a comparison
between seismic parameters and those derived from isochrone fitting is shown in
Figure \ref{fig2}. The level of agreement is reassuring, anticipating
that when {\it Gaia} parallaxes will be available, isochrone dating of turn-off
and subgiant stars will nicely complement with asteroseismology.

The breakthrough of asteroseismology, however, is the possibility of deriving
reliable stellar ages for red giant stars, where traditional isochrone fitting
fails. This is due to the fact that along the red giant branch isochrones with
vastly different ages can fit observational data such as effective
temperatures, metallicities, and surface gravities equally well within their
errors. However, once a star has evolved into a red giant, its age is determined
to good approximation by the
time spent in the hydrogen burning phase, and this is predominantly a function
of mass. Thus, seismic masses provide powerful age constraints. It should be
noted however that in case of mass-loss, the actual mass measured by
seismology will be somewhat smaller than the initial stellar mass (which sets
the evolutionary timescale), and uncertainties related to mass-loss might be
the main limiting factor to derive precise seismic ages for red giants.
Another advantage of using red giants stems from their intrinsic luminosities,
meaning that they can be easily used to probe distances up to a few kpc. 

Various schemes are now being developed to derive seismic ages across the HR
diagram, using different stellar models, and assessing the impact of different
input physics \citep[e.g.,][]{vsa13,vsa15,lg14,rgm14}. The first effort of
using seismic ages for the sake of Galactic studies has been put into place
with the Str\"omgren survey for Asteroseismology and Galactic Archaeology
(SAGA), which so far has derived classical and seismic stellar parameters for
giants in the {\it Kepler} field, covering Galactic latitudes from about
$8^{\circ}$ to $20^{\circ}$, at a fixed Galactic longitude, implying nearly
constant Galactocentric distances. This choice of geometry minimizes radial
variations, and makes the sample ideal to study the vertical structure of the
disc, thus allowing a direct measurement of the vertical age gradient
kinematically inferred from the GCS (see previous Section). SAGA builds on
the legacy of the
GCS by using Str\"omgren photometry, although it leverages on
asteroseismology. For a detailed accounting of the SAGA survey and its 
application to Galactic studies see \cite{c14a,csv15}.

The right hand panel in Figure \ref{fig1} shows the age-metallicity relation
derived
using red giants with seismic ages. This plot has been corrected to factor out
the main biases due to the complicated {\it Kepler} selection function, as
well as target selection effects. Similarly to what is seen in the solar
neighbourhood, also for the {\it Kepler} field there is a rather flat
age-metallicity relation. There is a somewhat larger metallicity spread,
although this is likely due to the lower precision of the Str\"omgren [Fe/H]
calibration adopted for giants in SAGA, compared to dwarfs and subgiants in
the GCS.
It should also be pointed out that sample selection excludes metal poor giants,
thus preventing us from tracing the early chemical enrichment, well visible
instead in the steep rise at $\sim 12$~Gyr for the Solar neighbourhood
(left hand panel).

\section{Selection functions and selection effects: preventing them, dealing
with them}

As we have already pointed out, despite being spatially limited, the GCS is
ideal for a number of Galactic studies because of its volume complete nature.
However, more often than not, this is not the case for other surveys. For
example, pencil
beam geometries are prone to various biases \citep[among which a correlation
  between Galactic heights and distances, see e.g.,][]{sac14}. 
In the case of the {\it Kepler} field, further biases stem from the fact that
the mission was not designed for population nor Galactic
studies, and thus it is not straightforward to assess whether stars with
different properties have been preferentially, or not, observed.

As it is often the case, before using any sample of stars for studies dealing
with Galactic structure, two questions must be answered. First, the {\it
  survey selection function} must the understood. In the case of {\it Kepler}
it means understanding to which extent asteroseismic giants are 
representative of the underlying population of giants in the field.
The best way of assessing this is by using a sample unbiased over a large
colour and magnitude range, built e.g., from a magnitude complete photometric
catalogue. More specifically, this means
benchmarking the asteroseismic sample against a (photometrically) unbiased
sample, to derive the colour and magnitude ranges
where the asteroseismic sample can be thought as randomly drawn from the field.
In the ideal case, a survey selection function should be clearly defined
before observations even begin, although contingent situations (known and
unknown unknowns, {\it sic}) might also play a role in modifying it. The
crucial point is that a survey selection function should always be defined upon
simple observables, such as colour and magnitude intervals. More generally,
one can also envisage doing this in a multi-dimensional space, using different
combinations of filters \citep[see e.g.,][for the sensitivity of bandpasses to
  various parameters]{b05,cv14}. Colour and magnitude intervals should be
sufficiently large to minimize boundary
effects, such as diffusion of stars in and out of the boundaries due to
observational errors. Reddening will also play a role, and again defining
sufficiently large intervals will minimize the impact of extinction-driven
colour and magnitude shifts across a selection function. Also, reddening
corrections should not be applied beforehand, or a selection function will
depend upon our (in)ability to estimate reddening at a given time. More
generally, while observations will stay unchanged, defining a selection
function upon variables which are not observed (including stellar parameters)
will make the selection itself subject to the unavoidable biases and
inaccuracies of those
variables. Further, future improvements in determining
certain parameters might clutter the original selection criteria, thus making
very hard --if at all possible-- to have control on them. On the contrary, a
clear and simple selection based on observables can always be reproduced, and
future improvements in determining stellar parameters, correcting for reddening,
predicting synthetic colours, etc\ldots can always be incorporated when forward
modelling observations. 

Once clear selection criteria are defined, we must still quantify the
probability that a star with certain stellar parameters will be observed. This
is due to {\it target selection effects}, i.e. how stellar parameters affect the
location of a star on the HR diagram, and thus how likely it is that a star
having certain parameters will pass our colour and magnitude cuts. To assess
these effects, a certain degree of modelling is needed. See e.g.,
\cite{csv15} for a discussion of different approaches to cope with the {\it
  target selection function} and {\it target selection effects} of seismic
targets in the {\it Kepler} field; an excellent and more general discussion can
be found in \cite{rb13}.

Looking ahead, the prospects of current and future asteroseismic missions for
Galactic studies are bright. To this purpose, now that {\it CoRoT} and
{\it Kepler} have demonstrated the potential of using red-giants for Galactic
studies \citep[e.g.,][]{miglio13a,chia15,mar15,csv15}, a better selection
function has been adopted for {\it K2} \citep{k2dennis}. The relevance of
Galactic science has also surged in future space-borne missions with
asteroseismic
capabilities, such as TESS, PLATO and WFIRST. Much before those, {\it Gaia}
astrometry will dramatically increase the volume for which we will know the
full stellar phase space information, and soon make possible to derive ages
for turn-off and subgiant stars across different Galactic components. The
mapping and dating of our Galaxy have just begun!

\acknowledgements
I thank the organizers of the conference for the invitation, the convivial
atmosphere created at the Physikzentrum Bad Honnef, and the financial support.

\bibliographystyle{an}

\begin{thebibliography}{}
\bibitem[{{Allende~Prieto} {et~al.}(2004){Allende~Prieto}, {Barklem},
  {Lambert}, \& {Cunha}}]{allende04:s4n}
{Allende~Prieto}, C., {Barklem}, P.~S., {Lambert}, D.~L., \& {Cunha}, K. 2004,
  \aap, 420, 183

\bibitem[{{Bensby} {et~al.}(2014){Bensby}, {Feltzing}, \& {Oey}}]{bfo14}
{Bensby}, T., {Feltzing}, S., \& {Oey}, M.~S. 2014, \aap, 562, A71

\bibitem[Bessell(2005)]{b05} Bessell, M.~S.\ 2005, \araa, 43, 293 

\bibitem[Binney et al.(2000)]{bi00} Binney, J., Dehnen, W., \& Bertelli, G.\ 2000, \mnras, 318, 658 
  
\bibitem[{{Bird} {et~al.}(2013){Bird}, {Kazantzidis}, {Weinberg}, {Guedes},
  {Callegari}, {Mayer}, \& {Madau}}]{bird13}
{Bird}, J.~C., {Kazantzidis}, S., {Weinberg}, D.~H., {et~al.} 2013, \apj, 773,
  43
\bibitem[Carlberg et al.(1985)]{car85} Carlberg, R.~G., 
Dawson, P.~C., Hsu, T., \& Vandenberg, D.~A.\ 1985, \apj, 294, 674 
  
\bibitem[{{Casagrande}(2015)}]{c15}
{Casagrande}, L. 2015, Astrophysics and Space Science Proceedings, 39, 61

\bibitem[{{Casagrande} \& {VandenBerg}(2014)}]{cv14}
{Casagrande}, L. \& {VandenBerg}, D.~A. 2014, \mnras, 444, 392

\bibitem[{{Casagrande} {et~al.}(2011){Casagrande}, {Sch{\"o}nrich}, {Asplund},
  {Cassisi}, {Ram{\'{\i}}rez}, {Mel{\'e}ndez}, {Bensby}, \& {Feltzing}}]{c11}
{Casagrande}, L., {Sch{\"o}nrich}, R., {Asplund}, M., {et~al.} 2011, \aap, 530,
  A138

\bibitem[{{Casagrande} {et~al.}(2014){Casagrande}, {Silva Aguirre}, {Stello},
  {Huber}, {Serenelli}, {Cassisi}, {Dotter}, {Milone}, {Hodgkin}, {Marino},
  {Lund}, {Pietrinferni}, {Asplund}, {Feltzing}, {Flynn}, {Grundahl}, {Nissen},
  {Sch{\"o}nrich}, {Schlesinger}, \& {Wang}}]{c14a}
{Casagrande}, L., {Silva Aguirre}, V., {Stello}, D., {et~al.} 2014, \apj, 787,
  110

\bibitem[Casagrande et al.(2016)]{csv15} Casagrande, L., 
Silva Aguirre, V., Schlesinger, K.~J., et al.\ 2016, \mnras, 455, 987 

\bibitem[{{Chaplin} {et~al.}(2014){Chaplin}, {Basu}, {Huber}, {Serenelli},
  {Casagrande}, {Silva Aguirre}, {Ball}, {Creevey}, {Gizon}, {Handberg},
  {Karoff}, {Lutz}, {Marques}, {Miglio}, {Stello}, {Suran}, {Pricopi},
  {Metcalfe}, {Monteiro}, {Molenda-{\.Z}akowicz}, {Appourchaux},
  {Christensen-Dalsgaard}, {Elsworth}, {Garc{\'{\i}}a}, {Houdek}, {Kjeldsen},
  {Bonanno}, {Campante}, {Corsaro}, {Gaulme}, {Hekker}, {Mathur}, {Mosser},
  {R{\'e}gulo}, \& {Salabert}}]{chap14}
{Chaplin}, W.~J., {Basu}, S., {Huber}, D., {et~al.} 2014, \apjs, 210, 1

\bibitem[{{Chaplin} \& {Miglio}(2013)}]{cm13}
{Chaplin}, W.~J. \& {Miglio}, A. 2013, \araa, 51, 353

\bibitem[{{Chiappini} {et~al.}(2001){Chiappini}, {Matteucci}, \&
  {Romano}}]{chia01}
{Chiappini}, C., {Matteucci}, F., \& {Romano}, D. 2001, \apj, 554, 1044

\bibitem[Chiappini et al.(2015)]{chia15} Chiappini, C., Anders, F., Rodrigues, T.~S., et al.\ 2015, \aap, 576, L12 

\bibitem[{{Coelho} {et~al.}(2015){Coelho}, {Chaplin}, {Basu}, {Serenelli},
  {Miglio}, \& {Reese}}]{coelho}
{Coelho}, H.~R., {Chaplin}, W.~J., {Basu}, S., {et~al.} 2015, \mnras, 451, 3011

\bibitem[{{De Cat} {et~al.}(2015){De Cat}, {Fu}, {Ren}, {Yang}, {Shi}, {Luo},
  {Yang}, {Wang}, {Zhang}, {Shi}, {Zhang}, {Dong}, {Catanzaro}, {Corbally},
  {Frasca}, {Gray}, {Molenda-Zakowicz}, {Uytterhoeven}, {Briquet}, {Bruntt},
  {Frandsen}, {Kiss}, {Kurtz}, {Marconi}, {Niemczura}, {Oestensen}, {Ripepi},
  {Smalley}, {Southworth}, {Szabo}, {Telting}, {Karoff}, {Silva Aguirre}, {Wu},
  {Hou}, {Jin}, \& {Zhou}}]{dc15}
{De Cat}, P., {Fu}, J.~N., {Ren}, A.~B., {et~al.} 2015, ArXiv e-prints

\bibitem[{{De Silva} {et~al.}(2015){De Silva}, {Freeman}, {Bland-Hawthorn},
  {Martell}, {de Boer}, {Asplund}, {Keller}, {Sharma}, {Zucker}, {Zwitter},
  {Anguiano}, {Bacigalupo}, {Bayliss}, {Beavis}, {Bergemann}, {Campbell},
  {Cannon}, {Carollo}, {Casagrande}, {Casey}, {Da Costa}, {D'Orazi}, {Dotter},
  {Duong}, {Heger}, {Ireland}, {Kafle}, {Kos}, {Lattanzio}, {Lewis}, {Lin},
  {Lind}, {Munari}, {Nataf}, {O'Toole}, {Parker}, {Reid}, {Schlesinger},
  {Sheinis}, {Simpson}, {Stello}, {Ting}, {Traven}, {Watson}, {Wittenmyer},
  {Yong}, \& {{\v Z}erjal}}]{ds15}
{De Silva}, G.~M., {Freeman}, K.~C., {Bland-Hawthorn}, J., {et~al.} 2015,
  \mnras, 449, 2604

\bibitem[{{Edvardsson} {et~al.}(1993){Edvardsson}, {Andersen}, {Gustafsson},
  {Lambert}, {Nissen}, \& {Tomkin}}]{edvardsson93}
{Edvardsson}, B., {Andersen}, J., {Gustafsson}, B., {et~al.} 1993, \aap, 275,
  101

\bibitem[{{Gliese}(1957)}]{gliese57}
{Gliese}, W. 1957, Astron.~Rechen-Institut, Heidelberg, 89 Seiten, 8, 1

\bibitem[Haywood et al.(2013)]{hay13} Haywood, M., Di Matteo, P., Lehnert, M.~D., Katz, D., \& G{\'o}mez, A.\ 2013, \aap, 560, A109 

\bibitem[{{Holmberg} {et~al.}(2007){Holmberg}, {Nordstr{\"o}m}, \&
  {Andersen}}]{holmberg07}
{Holmberg}, J., {Nordstr{\"o}m}, B., \& {Andersen}, J. 2007, \aap, 475, 519

\bibitem[{{Holmberg} {et~al.}(2009){Holmberg}, {Nordstr{\"o}m}, \&
  {Andersen}}]{holmberg09}
{Holmberg}, J., {Nordstr{\"o}m}, B., \& {Andersen}, J. 2009, \aap, 501, 941

\bibitem[{{Huber} {et~al.}(2012){Huber}, {Ireland}, {Bedding}, {Brand{\~a}o},
  {Piau}, {Maestro}, {White}, {Bruntt}, {Casagrande}, {Molenda-{\.Z}akowicz},
  {Silva Aguirre}, {Sousa}, {Barclay}, {Burke}, {Chaplin},
  {Christensen-Dalsgaard}, {Cunha}, {De Ridder}, {Farrington}, {Frasca},
  {Garc{\'{\i}}a}, {Gilliland}, {Goldfinger}, {Hekker}, {Kawaler}, {Kjeldsen},
  {McAlister}, {Metcalfe}, {Miglio}, {Monteiro}, {Pinsonneault}, {Schaefer},
  {Stello}, {Stumpe}, {Sturmann}, {Sturmann}, {ten Brummelaar}, {Thompson},
  {Turner}, \& {Uytterhoeven}}]{hub12}
{Huber}, D., {Ireland}, M.~J., {Bedding}, T.~R., {et~al.} 2012, \apj, 760, 32

\bibitem[Ivezi{\'c} et al.(2012)]{ibj12} Ivezi{\'c}, {\v Z}., Beers, T.~C., \& Juri{\'c}, M.\ 2012, \araa, 50, 251 

\bibitem[{{J{\o}rgensen} \& {Lindegren}(2005)}]{jl05}
{J{\o}rgensen}, B.~R. \& {Lindegren}, L. 2005, \aap, 436, 127

\bibitem[{{Just} \& {Jahrei{\ss}}(2010)}]{jj10}
{Just}, A. \& {Jahrei{\ss}}, H. 2010, \mnras, 402, 461

\bibitem[{{Kubryk} {et~al.}(2015){Kubryk}, {Prantzos}, \&
  {Athanassoula}}]{kpa15}
{Kubryk}, M., {Prantzos}, N., \& {Athanassoula}, E. 2015, \aap, 580, A126

\bibitem[{{Lebreton} \& {Goupil}(2014)}]{lg14}
{Lebreton}, Y. \& {Goupil}, M.~J. 2014, \aap, 569, A21

\bibitem[{{Lynden-Bell}(1975)}]{lb75}
{Lynden-Bell}, D. 1975, Vistas in Astronomy, 19, 299

\bibitem[Martig et al.(2015)]{mar15} Martig, M., Rix, H.-W., 
Aguirre, V.~S., et al.\ 2015, \mnras, 451, 2230 

\bibitem[{{Matteucci} \& {Francois}(1989)}]{mf89}
{Matteucci}, F. \& {Francois}, P. 1989, \mnras, 239, 885

\bibitem[{{Miglio} {et~al.}(2013{\natexlab{a}}){Miglio}, {Chiappini}, {Morel},
  {Barbieri}, {Chaplin}, {Girardi}, {Montalb{\'a}n}, {Noels}, {Valentini},
  {Mosser}, {Baudin}, {Casagrande}, {Fossati}, {Silva Aguirre}, \&
  {Baglin}}]{miglio13b}
{Miglio}, A., {Chiappini}, C., {Morel}, T., {et~al.} 2013{\natexlab{a}}, in
  European Physical Journal Web of Conferences, Vol.~43, European Physical
  Journal Web of Conferences, 3004

\bibitem[{{Miglio} {et~al.}(2013{\natexlab{b}}){Miglio}, {Chiappini}, {Morel},
  {Barbieri}, {Chaplin}, {Girardi}, {Montalb{\'a}n}, {Valentini}, {Mosser},
  {Baudin}, {Casagrande}, {Fossati}, {Aguirre}, \& {Baglin}}]{miglio13a}
{Miglio}, A., {Chiappini}, C., {Morel}, T., {et~al.} 2013{\natexlab{b}},
  \mnras, 429, 423

\bibitem[{{Minchev} {et~al.}(2013){Minchev}, {Chiappini}, \& {Martig}}]{min13}
{Minchev}, I., {Chiappini}, C., \& {Martig}, M. 2013, \aap, 558, A9

\bibitem[{{Nordstr{\"o}m} {et~al.}(2004){Nordstr{\"o}m}, {Mayor}, {Andersen},
  {Holmberg}, {Pont}, {J{\o}rgensen}, {Olsen}, {Udry}, \&
  {Mowlavi}}]{nordstrom04}
{Nordstr{\"o}m}, B., {Mayor}, M., {Andersen}, J., {et~al.} 2004, \aap, 418, 989

\bibitem[{{Perryman} {et~al.}(1997){Perryman}, {Lindegren}, {Kovalevsky},
  {Hoeg}, {Bastian}, {Bernacca}, {Cr{\'e}z{\'e}}, {Donati}, {Grenon}, {van
  Leeuwen}, {van der Marel}, {Mignard}, {Murray}, {Le Poole}, {Schrijver},
  {Turon}, {Arenou}, {Froeschl{\'e}}, \& {Petersen}}]{perryman97}
{Perryman}, M.~A.~C., {Lindegren}, L., {Kovalevsky}, J., {et~al.} 1997, \aap,
  323, L49

\bibitem[{{Pinsonneault} {et~al.}(2014){Pinsonneault}, {Elsworth}, {Epstein},
  {Hekker}, {M{\'e}sz{\'a}ros}, {Chaplin}, {Johnson}, {Garc{\'{\i}}a},
  {Holtzman}, {Mathur}, {Garc{\'{\i}}a P{\'e}rez}, {Silva Aguirre}, {Girardi},
  {Basu}, {Shetrone}, {Stello}, {Allende Prieto}, {An}, {Beck}, {Beers},
  {Bizyaev}, {Bloemen}, {Bovy}, {Cunha}, {De Ridder}, {Frinchaboy},
  {Garc{\'{\i}}a-Hern{\'a}ndez}, {Gilliland}, {Harding}, {Hearty}, {Huber},
  {Ivans}, {Kallinger}, {Majewski}, {Metcalfe}, {Miglio}, {Mosser}, {Muna},
  {Nidever}, {Schneider}, {Serenelli}, {Smith}, {Tayar}, {Zamora}, \&
  {Zasowski}}]{pin14}
{Pinsonneault}, M.~H., {Elsworth}, Y., {Epstein}, C., {et~al.} 2014, \apjs,
  215, 19

\bibitem[{{Pont} \& {Eyer}(2004)}]{pe04}
{Pont}, F. \& {Eyer}, L. 2004, \apj, 351, 487

\bibitem[{{Portinari} {et~al.}(1998){Portinari}, {Chiosi}, \&
  {Bressan}}]{pcb98}
{Portinari}, L., {Chiosi}, C., \& {Bressan}, A. 1998, \aap, 334, 505

\bibitem[{{Ram{\'{\i}}rez} {et~al.}(2007){Ram{\'{\i}}rez}, {Allende~Prieto}, \&
  {Lambert}}]{ramirez07}
{Ram{\'{\i}}rez}, I., {Allende~Prieto}, C., \& {Lambert}, D.~L. 2007, \aap,
  465, 271

\bibitem[{{Reddy} {et~al.}(2006){Reddy}, {Lambert}, \& {Allende
  Prieto}}]{reddy06}
{Reddy}, B.~E., {Lambert}, D.~L., \& {Allende Prieto}, C. 2006, \mnras, 367,
  1329

\bibitem[Rix \& Bovy(2013)]{rb13} Rix, H.-W., \& Bovy, J.\ 2013, \aapr, 21, 61 
  
\bibitem[{{Rodrigues} {et~al.}(2014){Rodrigues}, {Girardi}, {Miglio},
  {Bossini}, {Bovy}, {Epstein}, {Pinsonneault}, {Stello}, {Zasowski}, {Prieto},
  {Chaplin}, {Hekker}, {Johnson}, {M{\'e}sz{\'a}ros}, {Mosser}, {Anders},
  {Basu}, {Beers}, {Chiappini}, {da Costa}, {Elsworth}, {Garc{\'{\i}}a},
  {P{\'e}rez}, {Hearty}, {Maia}, {Majewski}, {Mathur}, {Montalb{\'a}n},
  {Nidever}, {Santiago}, {Schultheis}, {Serenelli}, \& {Shetrone}}]{rgm14}
{Rodrigues}, T.~S., {Girardi}, L., {Miglio}, A., {et~al.} 2014, \mnras, 445,
  2758

\bibitem[{{Ro{\v s}kar} {et~al.}(2008){Ro{\v s}kar}, {Debattista}, {Quinn},
  {Stinson}, \& {Wadsley}}]{roskar08}
{Ro{\v s}kar}, R., {Debattista}, V.~P., {Quinn}, T.~R., {Stinson}, G.~S., \&
  {Wadsley}, J. 2008, \apjl, 684, L79

\bibitem[{{Sch{\"o}nrich} {et~al.}(2014){Sch{\"o}nrich}, {Asplund}, \&
  {Casagrande}}]{sac14}
{Sch{\"o}nrich}, R., {Asplund}, M., \& {Casagrande}, L. 2014, \apj, 786, 7

\bibitem[{{Sch{\"o}nrich} \& {Binney}(2009)}]{sb09a}
{Sch{\"o}nrich}, R. \& {Binney}, J. 2009, \mnras, 396, 203

\bibitem[{Serenelli {et~al.}(2013)Serenelli, Bergemann, Ruchti, \&
  Casagrande}]{s13}
Serenelli, A.~M., Bergemann, M., Ruchti, G., \& Casagrande, L. 2013, MNRAS,
  429, 3645

\bibitem[{{Silva Aguirre} {et~al.}(2012){Silva Aguirre}, {Casagrande}, {Basu},
  {Campante}, {Chaplin}, {Huber}, {Miglio}, {Serenelli}, {Ballot}, {Bedding},
  {Christensen-Dalsgaard}, {Creevey}, {Elsworth}, {Garc{\'{\i}}a}, {Gilliland},
  {Hekker}, {Kjeldsen}, {Mathur}, {Metcalfe}, {Monteiro}, {Mosser},
  {Pinsonneault}, {Stello}, {Weiss}, {Tenenbaum}, {Twicken}, \&
  {Uddin}}]{vsa12}
{Silva Aguirre}, V., {Casagrande}, L., {Basu}, S., {et~al.} 2012, \apj, 757, 99
  
\bibitem[{{Silva Aguirre} {et~al.}(2013){Silva Aguirre}, {Basu}, {Brand{\~a}o},
  {Christensen-Dalsgaard}, {Deheuvels}, {Do{\u g}an}, {Metcalfe}, {Serenelli},
  {Ballot}, {Chaplin}, {Cunha}, {Weiss}, {Appourchaux}, {Casagrande},
  {Cassisi}, {Creevey}, {Garc{\'{\i}}a}, {Lebreton}, {Noels}, {Sousa},
  {Stello}, {White}, {Kawaler}, \& {Kjeldsen}}]{vsa13}
{Silva Aguirre}, V., {Basu}, S., {Brand{\~a}o}, I.~M., {et~al.} 2013, \apj,
  769, 141

\bibitem[{{Silva Aguirre} {et~al.}(2015){Silva Aguirre}, {Davies}, {Basu},
  {Christensen-Dalsgaard}, {Creevey}, {Metcalfe}, {Bedding}, {Casagrande},
  {Handberg}, {Lund}, {Nissen}, {Chaplin}, {Huber}, {Serenelli}, {Stello}, {Van
  Eylen}, {Campante}, {Elsworth}, {Gilliland}, {Hekker}, {Karoff}, {Kawaler},
  {Kjeldsen}, \& {Lundkvist}}]{vsa15}
{Silva Aguirre}, V., {Davies}, G.~R., {Basu}, S., {et~al.} 2015, \mnras, 452,
  2127

\bibitem[{{Stello} {et~al.}(2015){Stello}, {Huber}, {Sharma}, {Johnson},
  {Lund}, {Handberg}, {Buzasi}, {Silva Aguirre}, {Chaplin}, {Miglio},
  {Pinsonneault}, {Basu}, {Bedding}, {Bland-Hawthorn}, {Casagrande}, {Davies},
  {Elsworth}, {Garcia}, {Mathur}, {Di Mauro}, {Mosser}, {Schneider},
  {Serenelli}, \& {Valentini}}]{k2dennis}
{Stello}, D., {Huber}, D., {Sharma}, S., {et~al.} 2015, \apjl, 809, L3

\bibitem[{{Str\"omgren}(1987)}]{stromgren87}
{Str\"omgren}, B. 1987, in NATO ASIC Proc. 207: The Galaxy, ed. {G.~Gilmore \&
  B.~Carswell}, 229--246

\bibitem[{{Twarog}(1980)}]{twarog80}
{Twarog}, B.~A. 1980, \apj, 242, 242

\bibitem[{{van Leeuwen}(2007)}]{vanLeeuwen07}
{van Leeuwen}, F. 2007, \aap, 474, 653

\bibitem[{{Wallerstein}(1962)}]{wallerstein62}
{Wallerstein}, G. 1962, \apjs, 6, 407

\bibitem[Wielen(1977)]{w77} Wielen, R.\ 1977, \aap, 60, 263 

\bibitem[{White {et~al.}(2013)White, Huber, Maestro, Bedding, Ireland, Baron,
  Boyajian, Che, Monnier, Pope, Roettenbacher, Stello, Tuthill, Farrington,
  Goldfinger, Mcalister, Schaefer, Sturmann, Sturmann, Ten~Brummelaar, \&
  Turner}]{White:2013bu}
White, T.~R., Huber, D., Maestro, V., {et~al.} 2013, Monthly Notices of the
  Royal Astronomical Society, 433, 1262
\end{thebibliography}

\end{document}